\documentclass[a4paper]{eccomas_2022}
\usepackage{graphicx}
\usepackage{amsmath}
\usepackage{amsfonts}
\usepackage{amssymb}

\usepackage{graphicx}
\usepackage{amsmath}
\usepackage{amsfonts}
\usepackage{amssymb}

\usepackage{siunitx}
\usepackage{gensymb} 

\newcommand{\lowr}[1] {_{\mathrm{#1}}}
\newcommand{\upr}[1] {^{\mathrm{#1}}}

\usepackage[dvipsnames]{xcolor}

\usepackage{float}

\title{\uppercase{Approximating the full-field temperature evolution in 3D electronic systems from randomized ``Minecraft'' systems}}

\author{\uppercase{Monika Stipsitz}$^1$ AND \uppercase{H\`elios Sanchis-Alepuz}$^1$}

\heading{Monika Stipsitz and H\`elios Sanchis-Alepuz}

\address{$^{1}$ Silicon Austria Labs GmbH\\
Inffeldgasse 33, 8010 Graz, Austria\\
e-mail: \{monika.stipsitz, helios.sanchis-alepuz\}@silicon-austria.com
}


\abstract{Neural Networks as fast physics simulators have a large potential for many engineering design tasks. Prerequisites for a wide-spread application are an easy-to-use workflow for generating training datasets in a reasonable time, and the capability of the network to generalize to unseen systems. In contrast to most previous works where training systems are similar to the evaluation dataset, we propose to adapt the type of training system to the network architecture. Specifically, we apply a fully convolutional network and, thus, design 3D systems of randomly located voxels with randomly assigned physical properties. The idea is tested for the transient heat diffusion in electronic systems. Training only on random ``Minecraft'' systems, we obtain good generalization to electronic systems four times as large as the training systems (one-step prediction error of \SI{0.07}{\%} vs \SI{0.8}{\%}).}

\begin{document}

\section{INTRODUCTION}

The idea of using Neural Networks (NNs) as trainable physics simulators has seen much attention sparked by recent impressive results \cite{Sanchez-Gonzalez2020,Pfaff2020,Cranmer2020}. Once trained such a NN could predict physical properties much faster than any standard simulation method. Potential benefits can be seen in many fields of application, e.g.\ enabling large-scale, multi-domain, multi-objective design optimization, providing close to real-time feedback during a manual design process, as improved physics engine for computer games and in augmented reality applications. However, one obstacle are the large datasets required for the training process of such NNs. To be useful, for instance, in a design optimization process the time for generating the dataset should be small compared to the time a standard solver would spend in the optimization phase.

Regarding the data for the training of the NNs there are two main strategies: (1) Physics-inspired Neural Networks (PINNs) \cite{Karniadakis2021}, where the dataset is completely or partially replaced by an objective function based on the governing partial differential equation. This approach has proven successful for many applications \cite{Lim2022,Zhu2019} while convergence problems occurred for many others \cite{krishnapriyan2021characterizing,Wang2022a}. (2) A dataset is generated using a standard physics-based simulation method like Finite Element Method (FEM), Finite Volume Method, etc.\ \cite{Sanchez-Gonzalez2020,Pfaff2020}. These datasets are often similar or slightly smaller versions of the desired deployment systems.

\begin{figure}[h]
    \centering
    \includegraphics[trim=0cm 2.4cm 6cm 1cm, clip, width=0.9\textwidth]{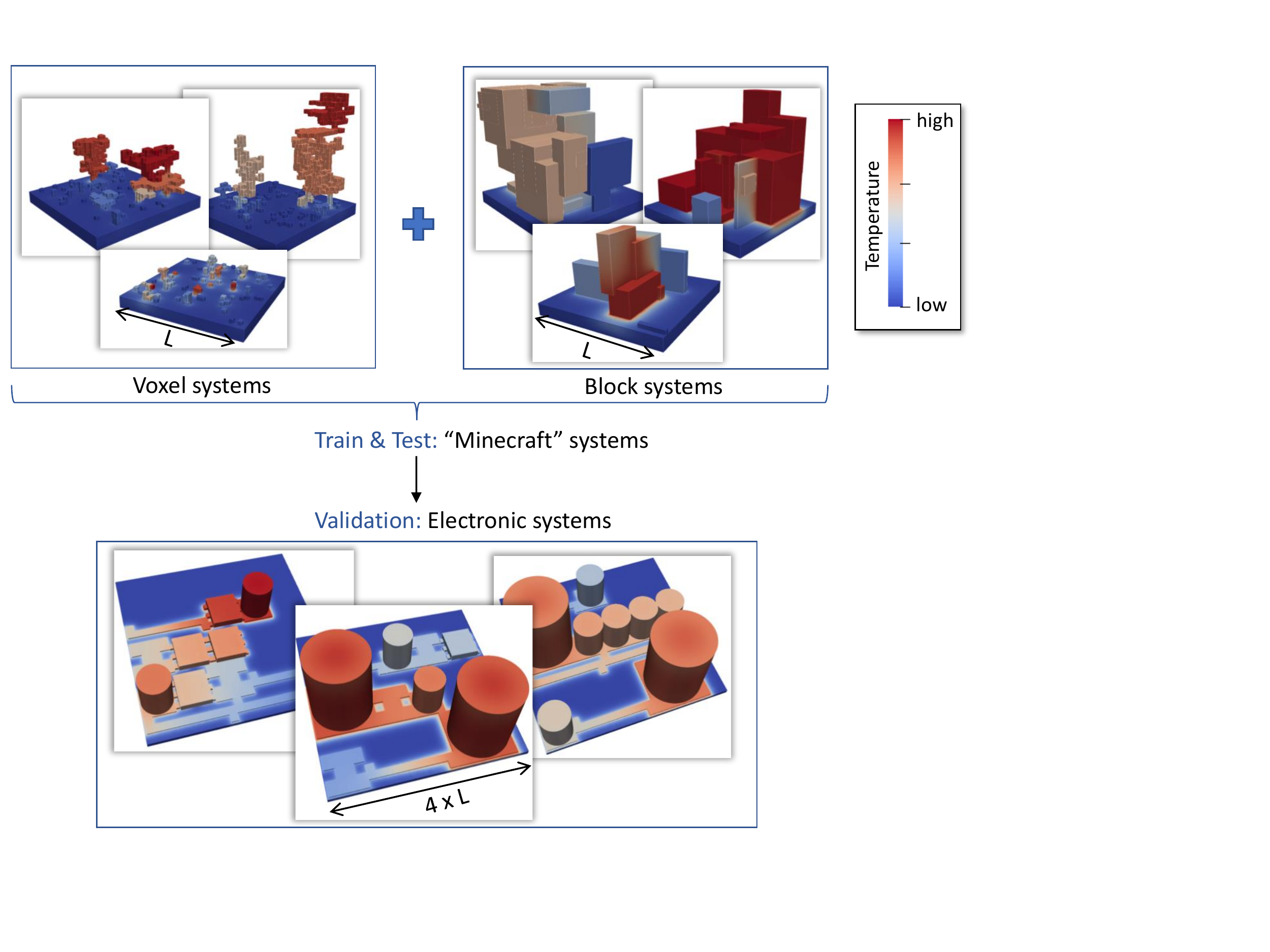} 
    \caption{Overview of the different system types: The NN is trained exclusively on random ``Minecraft'' systems of two different types (top). The generalization capabilities are validated on a set of electronic systems (bottom). The electronic systems are four times as large as the Minecraft systems.}\label{graphical_abstract}
\end{figure}

In a previous work, we observed that training a NN directly on a training dataset generated from the desired geometries might not be the best approach \cite{Stipsitz2022}. First, building systems from predefined components limits the geometrical diversity. The network tended to use the geometric shape of objects (e.g. small cylindrical) to predict their temperature rather than the physical properties (heat source, conductivity of surrounding region, etc.). Second, generating systems of similar size as the desired validation systems is time-consuming. Moreover, realistic systems in image representation take up most of the available memory of a standard graphics card. Thus, there is limited space for the network (and the gradients for the training process). Specifically, it was found that standard networks are too large. To capture the long-range correlations found in the heat propagation process a more intricate network architecture is necessary which, however, makes training the network more complicated.

Here, we suggest a different type of training dataset to overcome these problems. We apply a convolutional neural network (CNN) to obtain a fast approximation of the transient heat distribution in electronic systems. The major contributions are:
\begin{itemize}
    \item We investigate three-dimensional electronic systems with complex geometries consisting of different materials (see Fig. 1, bottom). The material properties span a wide range (e.g. copper, epoxy). The temperature distribution within the system is predicted.
    \item Instead of training directly on electronic systems, we exploit the fact that a fully convolutional network is used. We designed ``Minecraft'' systems consisting of randomly located blocks of randomly selected materials (see Fig. 1, top). The network trains only on these random systems but generalizes well to the electronic systems. The electronic systems are four times as large as the training systems. Due to memory limitations it would not be possible to train directly on the electronic systems.
\end{itemize}

\section{METHODS}

\begin{figure}[h]
        \includegraphics[width=\textwidth]{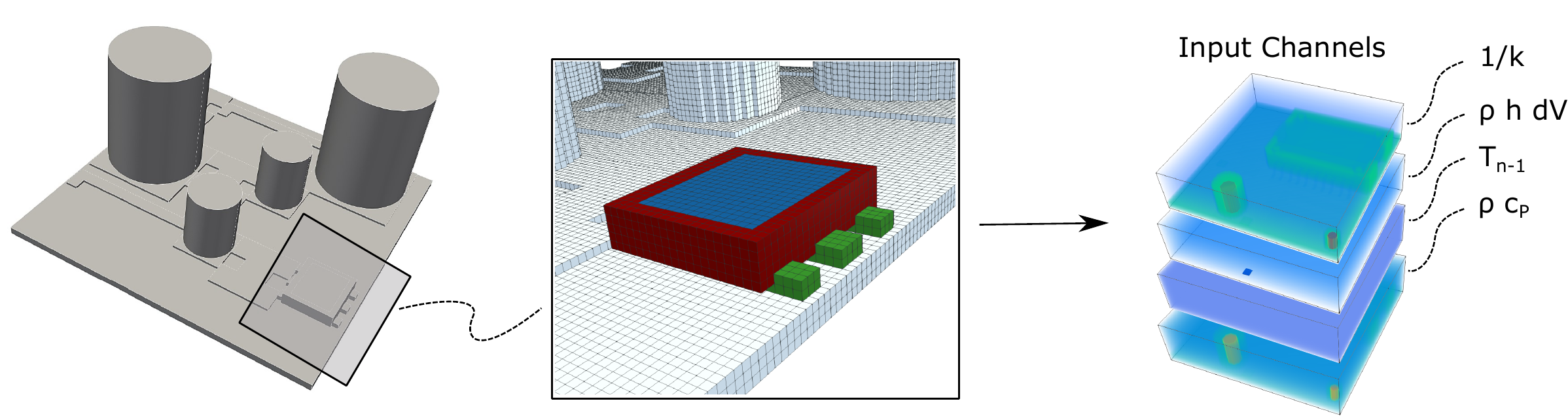}
        \caption{System representation: Starting from a CAD design (left), the components / the whole structure is split into voxels and combined to form an image of the original design (center). The input to the network consists of four channels, where each channel is a 3D image representing the distribution of the conductivity $k$, heat source power $\rho\, h\, \mathrm{d}V$, initial temperature $T_{n-1}$ or weighted capacity $\rho c\lowr{P}$ (right).}\label{image_generation}
    \end{figure}

    In this work we focus on the transient heat equation given by
    \begin{equation}
        \rho(\vec{x}) c_P(\vec{x})\frac{\partial T(\vec{x},t)}{\partial t}-\vec{\nabla}\cdot\left(k(\vec{x})\vec{\nabla} T(\vec{x},t)\right)- \rho(\vec{x}) h(\vec{x}) = 0~,
        \label{eq:HeatEq}
    \end{equation}
    where the temperature $T$, the heat source power per mass $h$, and the material properties density $\rho$, conductivity $k$ and heat capacity $c_P$. $\rho$, $h$, $k$ and $c_P$ depend on the position $\vec{x}$ but are constant over time $t$. All systems exhibit a bottom plate (see Fig.~\ref{image_generation}, left). The boundary conditions consist of a constant temperature $T\lowr{ext}$ at the bottom of the PCB plate, and a Neumann boundary condition on all other outer surfaces (heat transfer coefficient $\alpha = \SI{14}{W/(m^2 K)}$, external temperature $T\lowr{ext}$). Initially, the whole system is set to a uniform temperature equalling the temperature of the Dirichlet boundary condition ($T_0 = T\lowr{ext}$).

\begin{figure}[h]
        \includegraphics[width=\textwidth]{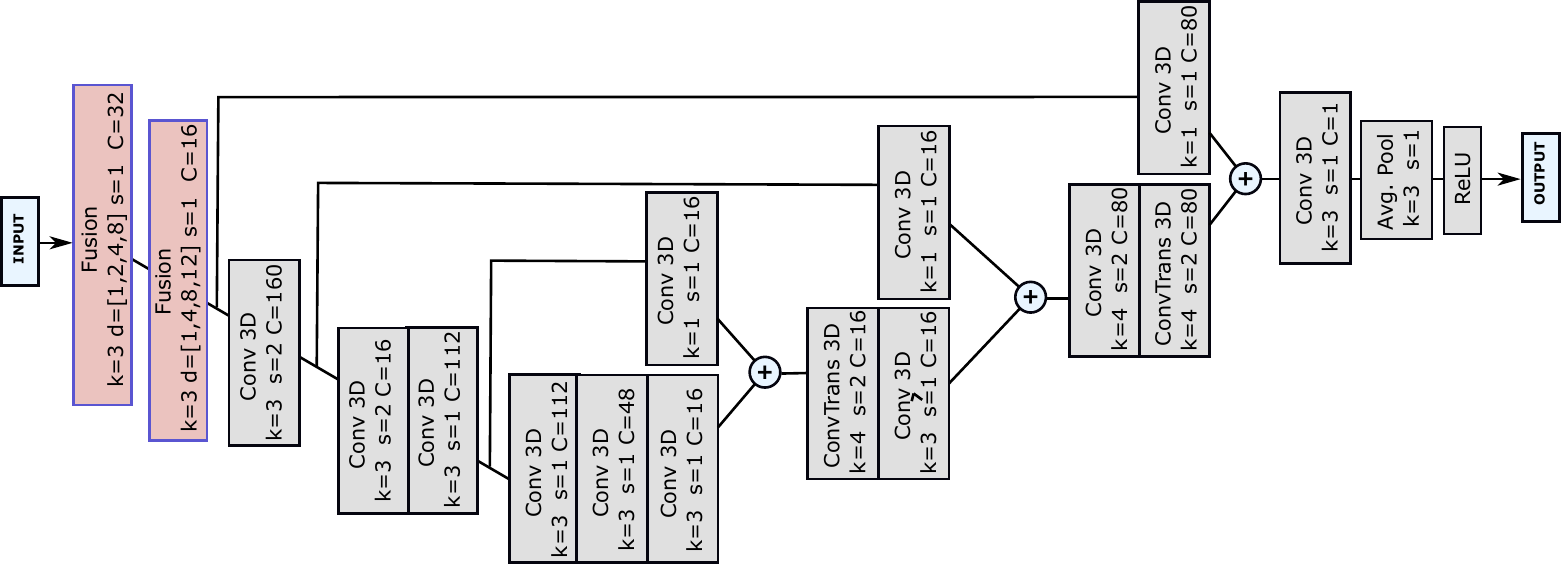}
        \caption{Fully convolutional architecture: Each ``Fusion'' block consists of 4 parallel 3D convolutions with different dilations $d$ and concatenated outputs. For each 3D convolutional layer (Conv) and transpose convolutional layer (ConvTrans) the kernel size $k$, stride $s$ and channel number $C$ of the output are given. Leaky ReLU is used as activation function where not stated otherwise. More details on the choice of architecture can be found in \cite{Stipsitz2022}.}\label{NN_architecture}
    \end{figure}

    A NN is constructed to predict the temperature distribution $T_n(\vec{x})$ given the temperature at the previous time step $T_{n-1}(\vec{x})$ with a time step of $\Delta t = \SI{2}{s}$. The temperature is represented as a 3D image, where each of the voxels is given the value of the temperature at the central points of a regular hexahedral grid with grid size \SI{0.2}{mm}.
    A fully convolutional architecture was chosen for the NN (Fig.~\ref{NN_architecture}, more details in \cite{Stipsitz2022}). The input to the network are four channels, with each channel a 3D image containing a physical property ($1/k$, $\rho h ~\mathrm{d}V$, $\rho c_P$), or the previous temperature distribution $T_{n-1}$ (see Fig.~\ref{image_generation}, right for an illustration of the channels). The physical properties are combined as they occur in Eq.~\ref{eq:HeatEq}, for instance, the second channel $\rho h ~\mathrm{d}V$ considers the integral heating power of a voxel with volume $\mathrm{d}V$. The channels are maximum-normalized using dataset statistics. The convolutional architecture naturally preserves translational equivariance \cite{Battaglia2018} and directly enables the network to be applied to larger images than it was trained on. Rotational equivariance was encouraged by data augmentation; the dataset is augmented by three 90 degree in-plane rotations of the images.

    Three different types of datasets are used within this work (compare Fig.~\ref{graphical_abstract}): Two types of random datasets (``Minecraft'' systems) which are used for training and testing, and a third dataset consisting of electronic systems which is used for validation only. The network will only see the Minecraft systems for training but should be able to generalize to the electronic systems.
    
    Since the network will see only patterns of voxels, an ideal training system would consist of completely randomly located voxels with randomly assigned values. If enough of these systems would be generated, all possible combinations of voxel values would occur on the level of the kernels of the network so that it should be able to generalize to any kind of system that can be defined as a stack of 3D images.

    In practice there are, however, a number of limitations on how the training systems can be designed. First, in order that FEM simulations can be performed the system has to be uniquely connected. Thus, we decided to grow all systems on a PCB plate. Second, to randomly create thousands of systems an automatized workflow is necessary. Specifically, FreeCAD and python were used to generate the systems. However, it was found that FreeCAD was not well accustomed to process multiple thousands of independent components. Third, an immense number of systems would be needed to represent all possible combinations. It is, thus, more reasonable to generate systems better adjusted to the desired validation systems.

    \textbf{Training dataset} The Minecraft systems should capture the geometric features of electronic systems. The electronic systems contain small details, e.g. copper tracks or legs, in the order of individual voxels, as well as larger components (e.g. molds), and parts (e.g. dies) completely enclosed in other components. Two types of random systems were generated to cover this variety of geometrical shapes (Fig.~\ref{graphical_abstract}): ``voxel systems'' constructed of individual voxels where each one is assigned a different random material, and ``block systems'' made up of larger boxes leading to larger regions with the same material properties. To increase the variation in geometrical shapes, the block systems were generated from randomly sized 3D blocks which were randomly placed on the PCB. The blocks could be overlapping, in which case the structure was decomposed into multiple non-overlapping segments.
    
    All random systems were generated with a size of \SI{6.4}{mm} per dimension (a quarter of the size of the electronic systems in the validation dataset). A voxel size of \SI{0.2}{mm} was chosen. Each voxel/block is assigned a homogeneous material. Thus, the voxel size limits the minimal size at which structures are resolved.

    \textbf{Validation dataset} The electronic systems had a size of \SI{25.6}{mm} per dimension. They were generated by randomly placing pre-designed components (one IC, one large and one small capacitor, and copper layers of different shapes and sizes) on a PCB. Most of these components have an internal structure, e.g. the large IC consists of a die, mold base, thermal pad, legs and copper layers.
        
    Finally, the training data for the supervised learning process was generated using an automatized workflow employing the open-source packages FreeCAD for system construction, gmsh for meshing and ElmerFem to solve for the temperature distributions:
    One of the materials in Tab.~\ref{material} was assigned randomly to each component of a system. (For the electronic systems the materials were not assigned randomly but the composition of each component was fixed, e.g. legs of copper, mold of epoxy.)
    \begin{table}
        \centering
        \caption{Average material properties: The actual values are chosen randomly per component from a range of $[0.75 \mathrm{Avg}, 1.25 \mathrm{Avg}]$.}\label{material}
        \begin{tabular}{lllll}
        Property & Avg. $k$  & Avg. $c_P$  & Avg. $\rho$  \\
        Unit    &  $(W / (m K))$ & $(J / (kg K)$ & $(kg / m^3)$ \\ \hline
        Silicon & 148 & 705 & 2330 \\
        Copper & 384 & 385 & 8930 \\
        Epoxy & 0.881 & 952 & 1682 \\
        $\mathrm{FR}_4$ & 0.25 & 1200 & 1900 \\
        $\mathrm{Al}_2 \mathrm{O}_3$ & 35 & 880 & 3890 \\
        Aluminium & 148 & 128 & 1930 \\
        \end{tabular}
    \end{table}
    The uniform initial temperature was chosen randomly from the range $[-40, 200]$ \degree C. Heat sources were assigned randomly to some of the components. The temperature was extracted at the central points of a hexahedral grid with element sizes of \SI{0.2}{mm} $\times$ \SI{0.2}{mm} $\times$ \SI{0.2}{mm} and a temporal discretization of $\Delta t = \SI{2}{s}$. Thus, the temperature at a given time step was represented as a 3D image of similar dimensions as the input channels (compare Fig.~\ref{image_generation}).  Unphysically hot systems with $T\lowr{max} > \SI{250}{K}$ were discarded based on a steady-state simulation. For the transient simulations, the systems started in a uniform temperature state and were heated up until close to the steady-state (after 15 time steps), from which four time-steps were randomly sampled for the datasets. In total, 2440 valid Minecraft systems were generated (1129 voxel systems and 1311 block systems). A split of \SI{75}{\%} was used for the train/test set.

\subsection{Impact of training dataset type}\label{Sec:system_type}

    To evaluate the generalization capabilities of the different types of training datasets, the network was trained with three different Minecraft datasets: (1) only voxel systems, (2) only block systems, (3) a combined dataset of comparable size as (1) and (2) to enable a fair comparison. All training procedures achieved comparable accuracy on the corresponding test datasets. After training, the network was evaluated on the electronic systems to evaluate the generalizability.

\begin{figure}[h]
        \centering
            \includegraphics[width=\textwidth]{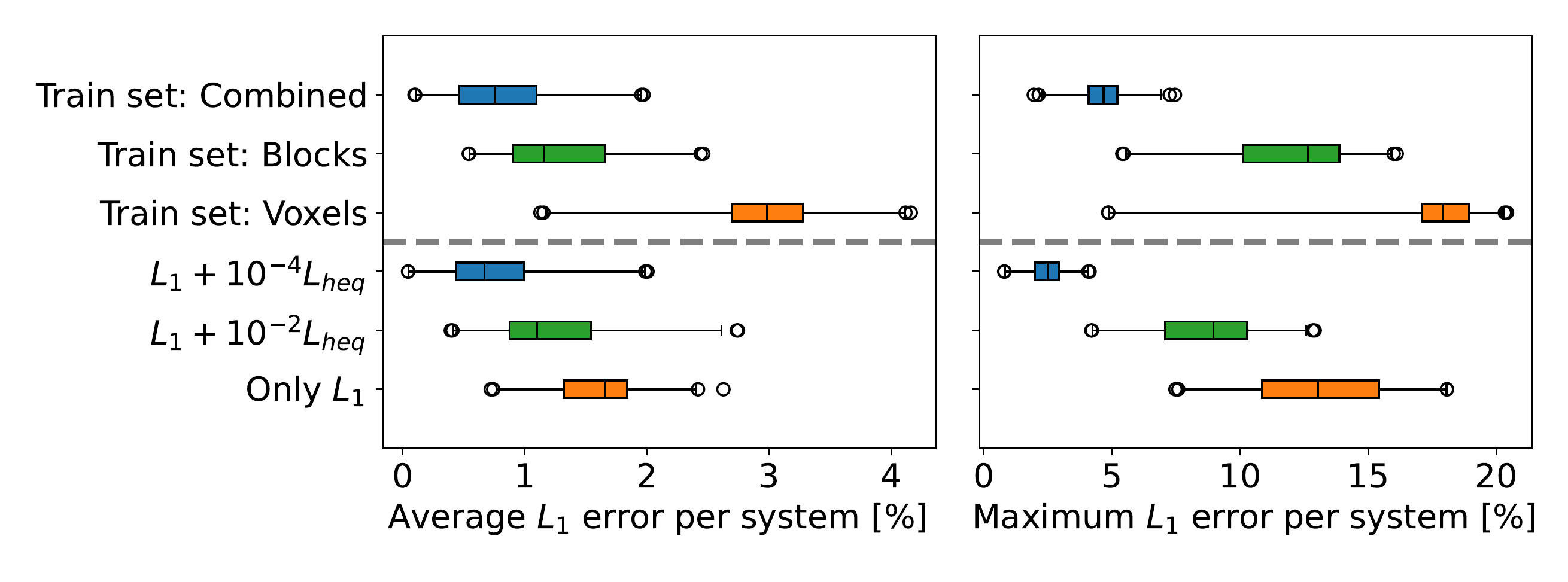}\\
        \caption{Box plot of the $L_1$ error (left: average, right: maximum per system) evaluated for the electronic systems. Rows 1-3 show the impact of different types of random systems as training datasets. Rows 4-6 evaluate the effect of different objective function compositions on the generalization capabilities. For more information see Secs.~\ref{Sec:system_type} and \ref{Sec:loss}, respectively.}\label{boxplot_error}
    \end{figure}

    Fig.~\ref{boxplot_error} reports the average and maximum $L_1$ error for the validation dataset evaluated per system. The best $L_1$ error was achieved with the combined dataset (see Fig.~\ref{boxplot_error}, rows 1-3), and also regarding the maximum error found in the systems the combined dataset was clearly superior to both the voxel and the block training datasets. The maximum $L_1$ error was nearly \SI{10}{\%} larger when trained with only voxel or only block systems (Fig.~\ref{boxplot_error}, right). The slightly worse errors in the combined dataset compared to the final trained network indicate that the NN can profit from the larger dataset (compare  ``Train set: Combined'' and  ``$L_1 + 10^{-4} L\lowr{heq}$'' in Fig.~\ref{boxplot_error}).

\begin{figure}[h]
        \centering
        \includegraphics[trim=0cm 0cm 0cm 12.5cm, clip,width=\textwidth]{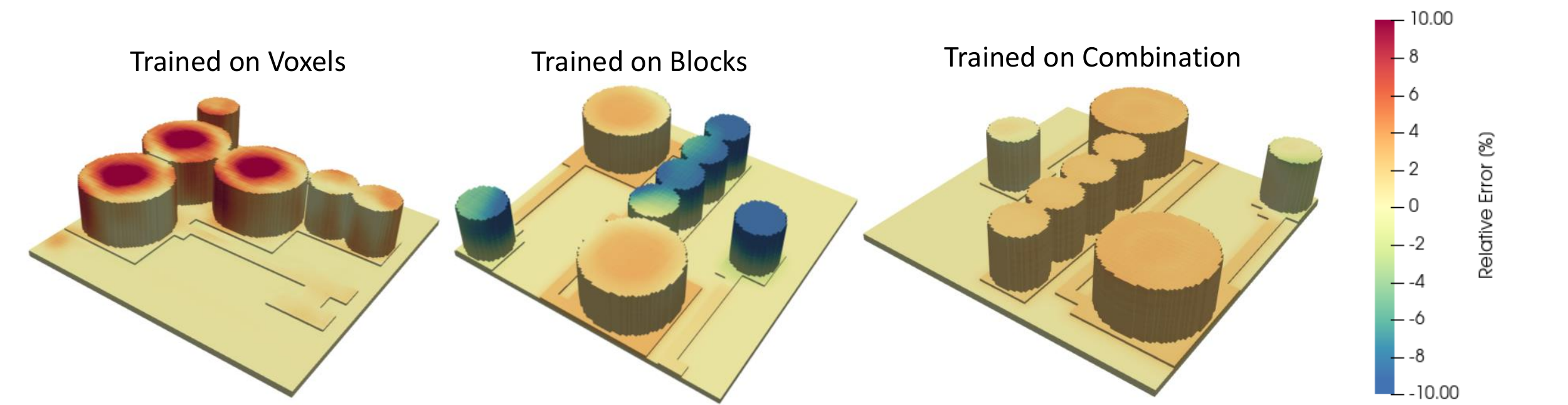}\\
        \caption{Relative error in the electronic system with the worst average $L_1$ error. The image shows only the lower half of the systems so that errors within the structures are visible. The evaluations were performed with three differently trained networks, once training using only the voxel systems (left), once only the block systems (center) and once the combined dataset of the same size (right). Note, that for visualization purposes the color scale for the error is capped at $\pm$ \SI{10}{\%}.}\label{local_dataset_type}
    \end{figure}

    Qualitatively, it is informative to look at the worst-case system, i.e.\ the system with the largest $L_1$ error (Fig.~\ref{local_dataset_type}). The idea of using two types of random systems was that the voxel systems would teach the network the importance of small details in the size of individual voxels while the block systems would help it to learn the behaviour of larger heat generating blocks. When trained with the voxel systems, it is clearly notable that the network is not used to larger areas containing heat sources, leading it to predict much hotter cores of the capacitors (Fig.~\ref{local_dataset_type}, left). However, the role of the block systems is less obvious (Fig.~\ref{local_dataset_type}, center). Although the prediction of larger heat generating components improved compared to the voxel systems and the copper layers show slightly higher errors, large errors are located at intermediately sized structures. Nevertheless, the combined dataset leads to smaller, more uniform errors (Fig.~\ref{local_dataset_type}, right).

\subsection{Optimization of learning scheme}\label{Sec:loss}

    Two types of objective functions were evaluated:
    \begin{itemize}
        \item A normalized $L_1$ loss given by $L_1 = \mathrm{Average}\left(\frac{T\upr{FEM} - T\upr{NN}}{T\upr{FEM}}\right)$, where $T\upr{FEM}$ and $T\upr{NN}$ are the baseline temperature map obtained from a FEM simulation and the predicted temperature map from the NN, respectively.
        \item A physics-based loss $L\lowr{heq}$ defined via an integrated and discretized version of the heat equation (Eq.~\ref{eq:HeatEq}) inspired by \cite{Ren2021}. The differential operators were evaluated for the same coarse temporal and spatial discretization used for the dataset. The $L\lowr{heq}$ was computed as the sum of the absolute values of this local residual. Due to the coarse discretization, $L\lowr{heq}$ evaluated for the FEM solutions was non-negligible; on average $L\lowr{heq}(T\upr{FEM}) = 1.95$.
    \end{itemize}
    Both objective functions were evaluated only for the image regions containing components.
    
    Different combinations of these two losses were tested. The network was trained on the Minecraft systems, and afterwards the generalization capabilities for the electronic systems was evaluated (Fig.~\ref{boxplot_error}, rows 4-6): 
    Training with $L_1$ loss, and training with $L_1$ loss plus a small physics-based regularization (with scaling of $10^{-2}$ or $10^{-4}$) led to similar average and maximum $L_1$ error distributions for the Minecraft systems (average $L_1$ error of \SI{0.07}{\%}, \SI{0.08}{\%} and  \SI{0.07}{\%}, respectively). In contrast, training only with $L\lowr{heq}$ led to much higher errors even on the Minecraft systems ($L_1 = \SI{10}{\%}$) and bad generalization to the electronic systems. This was expected, since the definition of $L\lowr{heq}$ relies on a very coarse discretization. However, adding $L\lowr{heq}$ as a physics-based regularization term to the $L_1$ error significantly improved the generalization capabilities compared to the $L_1$ loss only. It was previously argued that physics-based regularization terms improve learning because they add non-locality to the loss \cite{Chen2022}. For instance, they could encourage spatial smoothness often found in physical systems. However, it was surprising that even a bad approximation of the physical laws can enhance the learning process.

    The results did not improve when random noise was added to the input channels as suggested in \cite{Sanchez-Gonzalez2020}.

\section{RESULTS}\label{Sec:results}

    The results reported in this section are based on the training setup labelled ``$L_1 + 10^{-4}L\lowr{heq}$'' in Fig.~\ref{boxplot_error}. As training dataset a combination of the voxel and block type random systems were used. All results given for the Minecraft dataset refer to the test part of the dataset.

\subsection{One-step prediction}

\begin{figure}[h]
        \centering
        \includegraphics[trim=2cm 3.4cm 2cm 0.5cm, clip,width=\textwidth]{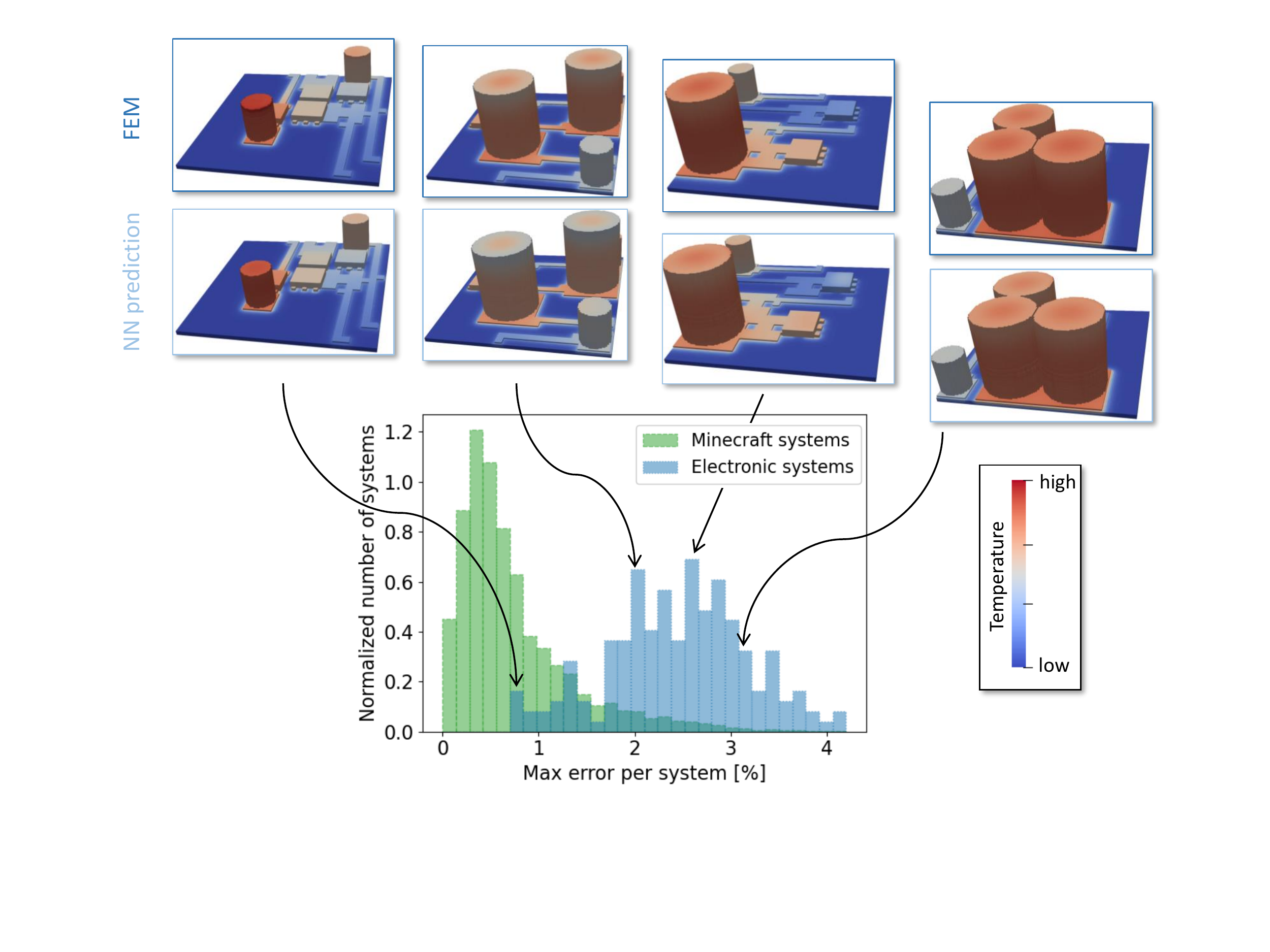}
        \caption{Histogram of the maximum $L_1$ error per system and selected temperature distribution predicted by the NN (bottom) and baseline from the FEM simulation (top).}\label{local_error}
    \end{figure}

    The network was trained on individual time steps only; given the temperature distribution at the current time step it was asked to predict the temperature distribution at the next time step.
    The mean $L_1$ error for the one-step prediction on the Minecraft dataset was \SI{0.07}{\%}. For the one-step prediction reasonably good generalization to the electronic systems was achieved (average $L_1$ error of \SI{0.8}{\%}) taking into account that the electronic systems are four times as large as the training systems. In more detail, the worst electronic system had an average error of $\SI{2}{\%}$ while the worst average error for Minecraft systems was \SI{0.9}{\%}. The maximum error for the electronic systems is only slightly larger than for the Minecraft systems (see histogram of the maximum errors per system in Fig.~\ref{local_error}).
    
    Qualitatively, the predicted temperature distributions compare well with the baseline from the FEM simulations (Fig.~\ref{local_error}). In some systems the predicted temperature at the surface is slightly too low. This is most probably caused by setting the background to zero in the input channels. Additionally, larger structures like the large capacitor show slight non-uniformities in the predicted temperature distributions.

\subsection{Multi-step prediction}\label{Sec:unrolling}

    After training on single-steps the NN was applied to predict the whole time series from a uniform temperature distribution until close to steady-state after 15 time steps. For this, the output of the NN was repeatedly fed back into the network. For the Minecraft test dataset the average $L_1$ error accumulates to \SI{0.47}{\%} $\pm$ \SI{0.3}{\%} after 15 time steps. The $L\lowr{heq}$ based regularization term remains constant over the unrolling steps ($(0.018 \pm 0.0008)\times 10^{-4}$). Visually, the temperature distribution in the Minecraft systems after 15 unrolling steps look still similar to the baseline FEM simulations.
    
\begin{figure}[h]
        \centering
        \includegraphics[trim=0cm 3.5cm 0cm 7cm, clip,width=\textwidth]{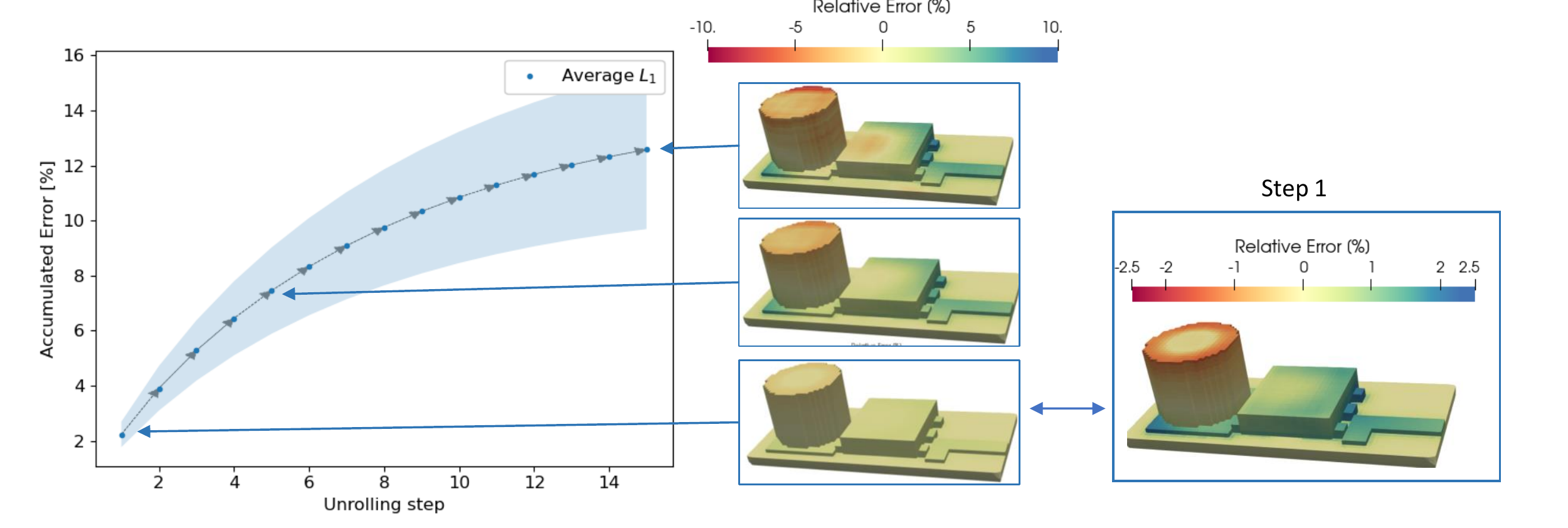}
        \caption{The accumulated error (average and variance region) is obtained by starting from an initially homogeneous temperature state, and feeding the predicted temperature distribution repeatedly back to the network during evaluation (left). Training was performed on single time steps only. The center and right show selected sections of a system after one, five and 15 time steps.}\label{chip_per_step}
    \end{figure}

    For the electronic systems the accumulated error is still bounded, although, much larger than for the Minecraft systems (Fig.~\ref{chip_per_step}). After one step the error is mostly distributed on the surface of the components (Fig.~\ref{chip_per_step}, right) and spreads from there with increasing number of steps. Arguably, for the application in electronic design the temperature at the center of components is of most interest since over-heating may occur here.

    It can be seen that the error accumulated in the first couple of steps is higher than for later steps. This could indicate a bias in the training dataset. To construct the Minecraft dataset the FEM simulations were sampled at 15 time steps. The dataset was then constructed by randomly sampling four of these one-step predictions. Step 15 was chosen to be close to the steady-state distribution. This might lead to an over-representation of small temperature changes per time step. This result indicates that the training could in future improve by non-uniform selection of training data from different initial time steps.

\subsection{Performance}\label{Sec:performance}

    The time for generating the training dataset was significantly reduced with the random dataset compared to a training dataset of electronic systems. The time for one of the transient FEM simulation was significantly reduced from on average \SI{5200}{s} for an electronic system to about \SI{800}{s} for a Minecraft type system (on a single thread of an Intel Xeon W-2145 CPU). The training was completed after 400 epochs \'a \SI{620}{s} (on an NVIDIA Titan RTX GPU). During deployment the trained network can predict the 3D temperate distribution for a batch of 40 electronic systems in \SI{70}{ms} including the time to move the systems to the GPU memory.
    
    Also, the smaller system sizes of the Minecraft dataset enable the training of reasonable large NNs (such as the one presented in this work) on a standard GPU. Using the electronic systems directly the same network would not be trainable because of the limited memory available on the GPU.


\section{DISCUSSION \& CONCLUSIONS}

    The task that we pose to the NN is challenging: The training systems are only a quarter of the size of the electronic systems. Thus, longer range correlations than the size of the training systems are never seen by the network. The spatial and temporal discretization is much coarser than would be applicable for a standard numerical method. The material properties span multiple orders of magnitude (compare Tab.~\ref{material}). The electronic systems consist of small details (like copper layers or legs which can be as small as one voxel) and larger homogeneous blocks. Additionally, the network is employed for multi-step prediction on the electronic systems although it was trained on one-step predictions of systems with random geometry only. Nevertheless, the network generalizes well to the electronic systems for the one-step predictions and leads to a bounded error in the unrolling.

    To summarize, there are obvious advantages in using a randomly generated dataset for training. The smaller system sizes lead to faster simulation times to obtain the ``true'' temperature distributions of the training systems (Sec.~\ref{Sec:performance}). In comparison with a previous work \cite{Stipsitz2022}, we found that the higher geometrical variability of the Minecraft systems makes the network less likely to lock to pattern recognition instead of using the physical properties. Additionally, it should better generalize to a wider range of different geometries than when trained on the goal systems only, as was shown for the electronic systems. Thus, no retraining should be necessary when new components / geometrical variations are added later on.

    One clear limitation of the type of training dataset presented here is that generating a good random dataset representing all the desired   features of the physical equation and desired geometry is a challenge on its own. For instance, systems that consist of randomly located voxels of random material properties and with heat sources assigned to some random voxels can exhibit unphysically hot locations. This can occur, for instance, if a heat source was placed among a mostly isolating region. Also, in order that the NN can generalize to the electronic systems the input and output values should be of the same range. In the case presented here, this means that all material properties, the heat source power per voxel as well as the overall temperature range and the temperature increase per time step should be of the same order.

    Representing the systems as images has clear advantages for the type of application we have in mind: In a manual design process, a designer could move components and the NN could return a near-instantaneous feedback on the thermal distribution. This is possible because the components can be pre-voxelized so that generating the input channels for a new design takes only fractions of a second. For this type of application a temperature feedback at coarse grid points would be sufficient. With other data representations, like meshes, the geometry could be much closer represented but moving components would require re-meshing.


    In this work, we presented a random dataset for training a fully convolutional network to predict the transient 3D temperature distribution in electronic systems. The NN was trained on the randomly generated geometries only but generalized well to electronic systems which were four times as large. The network was tested with a set of electronic systems with differently shaped components, different placement of those components, different material properties and different initial temperatures. Two points proved decisive to obtain good generalization to the electronic systems: First, two types of randomized datasets were employed to cover most geometrical features found in the electronic systems. Second, the network did not train well based only on the $L_1$ error but training and generalization capabilities were significantly improved when a physics-based regularization term was added to the objective function.

\section*{Acknowledgment}

This work has been supported by Silicon Austria  Labs GmbH (SAL), owned by the Republic of Austria, the Styrian Business Promotion Agency (SFG), the federal state of Carinthia, the Upper Austrian Research (UAR), and the Austrian Association for the Electric and Electronics Industry (FEEI).

\small

\bibliographystyle{ieeetr} 
\bibliography{fempap_biblio_v2}

\end{document}